\documentclass[copyright,creativecommons]{eptcs}
\providecommand{\event}{FMAS 2026} 

\usepackage{iftex}

\ifpdf
  \usepackage{underscore}         
  \usepackage[T1]{fontenc}        
\else
  \usepackage{breakurl}           
\fi

\usepackage{tikz}
\usetikzlibrary{arrows.meta,positioning,backgrounds,fit,calc,shapes.geometric}
\usepackage{graphicx}
\usepackage{subcaption}
\usepackage{amsmath}

\usepackage{placeins}

\usepackage{multirow}
\usepackage{siunitx}
\usepackage{float}
\usepackage{fancyvrb}                                     
\title{RyBench: Runtime Verification Benchmarking Framework over Publish/Subscribe Networks}

\author{Mehmet Hilmi Dündar
\institute{Department of Computer Engineering\\
Boğaziçi University\\
Istanbul, Turkey}
\email{mehmet.dundar@std.bogazici.edu.tr}
\and
Doğan Ulus
\institute{Department of Computer Engineering\\
Boğaziçi University\\
Istanbul, Turkey}
\email{dogan.ulus@bogazici.edu.tr}
}
\def\titlerunning{RyBench: RV Benchmarking Framework}
\def\authorrunning{M.H. Dündar \& D. Ulus}

\begin{document}
\maketitle

\begin{abstract}
Autonomous systems built as distributed, node-based systems communicate via publish/subscribe middleware.
Runtime Verification (RV) tools are used to monitor those systems against formal specifications, receiving monitored events through the pub/sub channel.
An RV tool's practical value depends on its monitoring throughput capacity and verdict latency, besides its correctness, so a systematic benchmarking setup is needed to evaluate how efficiently these tools run.
This paper presents RyBench, a pub/sub-based benchmarking framework that acts as a synthetic event publisher in place of a monitored system, providing a common benchmarking setup.
RyBench separates benchmark generation, serialization, and transport into distinct components, enabling the combination of different event-generation strategies, data formats, and transport layers.
We also evaluate the framework's throughput capacity across these configurations using Open Simulation Interface (OSI) schemas and Timescales properties, and illustrate the end-to-end benchmarking flow through an application to Reelay and an integration with MonPoly.
\end{abstract}

\section{Introduction}
\label{sec:introduction}


Autonomous systems such as robots, self-driving vehicles, and other cyber-physical platforms are expected to operate in continuous interaction with their environment, responding to dynamic conditions as \textit{reactive systems}.
The correctness of reactive systems depends on the temporal ordering and consistency of responses to external inputs.
\textit{Runtime Verification (RV)} aims to check the correctness of those systems by deriving monitors from formal specifications.
During execution, these monitors observe incoming events and produce verdicts indicating whether the observed execution satisfies, violates, or remains inconclusive.

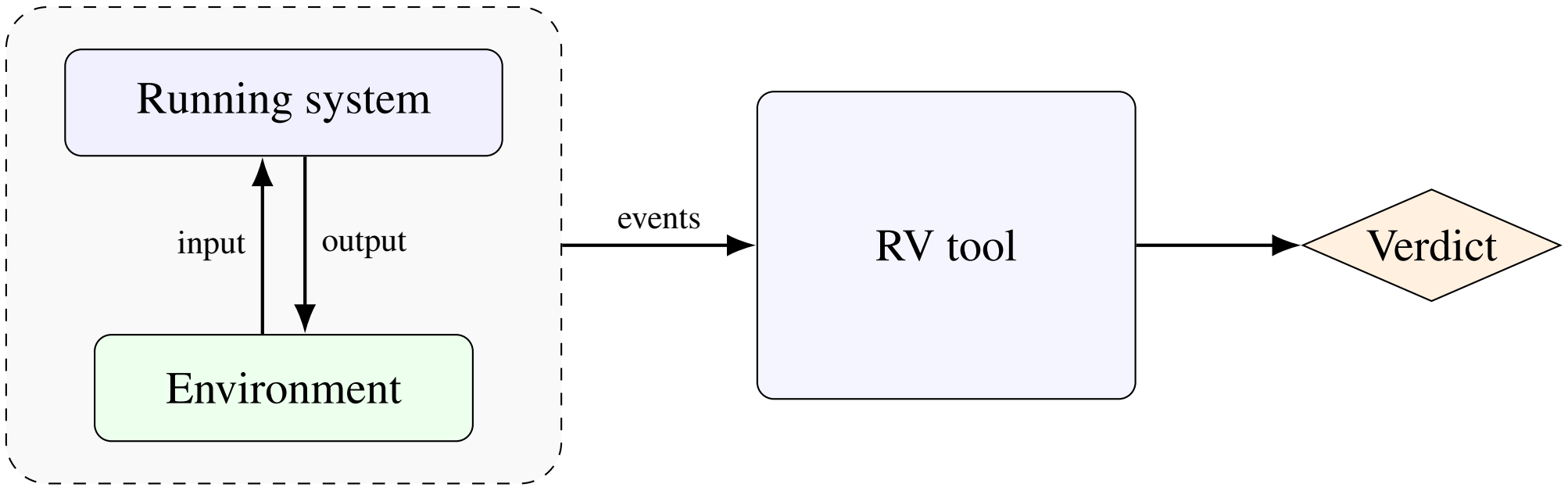
\begin{figure}[h]
\centering
\begin{tikzpicture}[
  node distance=0.5cm,
  block/.style={draw, rounded corners, align=center, minimum width=3.7cm, minimum height=0.9cm, fill=blue!6},
  env/.style={draw, rounded corners, align=center, minimum width=3.2cm, minimum height=0.9cm, fill=green!7},
  monitorbox/.style={draw, rounded corners, align=center, minimum width=3.2cm, minimum height=2.6cm, fill=blue!4},
  verdict/.style={draw, diamond, aspect=2.3, align=center, inner sep=2pt, fill=orange!12},
  contextbox/.style={draw, dashed, rounded corners=10pt, fill=gray!5, inner xsep=14pt, inner ysep=10pt},
  arrow/.style={-{Latex[length=2.5mm]}, thick}
]
\node[block] (system) {Running system};
\node[env, below=1.5cm of system] (environment) {Environment};
\begin{pgfonlayer}{background}
\node[contextbox, fit=(system)(environment)] (context) {};
\end{pgfonlayer}
\node[monitorbox, right=4cm of $(system)!0.5!(environment)$] (monitor) {RV tool};
\node[verdict, right=1.4cm of monitor] (verdict) {Verdict};
\draw[arrow] ([xshift=-0.18cm]environment.north) -- node[left, font=\scriptsize] {input} ([xshift=-0.18cm]system.south);
\draw[arrow] ([xshift=0.18cm]system.south) -- node[right, font=\scriptsize] {output} ([xshift=0.18cm]environment.north);
\draw[arrow] (context.east) -- node[above, font=\scriptsize] {events} (monitor.west);
\draw[arrow] (monitor.east) -- (verdict.west);
\end{tikzpicture}
\caption{Runtime verification monitoring.}
\label{fig:rv-monitor}
\end{figure}

As illustrated in Figure~\ref{fig:rv-monitor}, the running system interacts with its environment through inputs and outputs. At the same time, the RV tool consumes the resulting stream of events from outside the system boundary and produces a verdict as its output.
Formally, this event stream constitutes a trace $\sigma$, and runtime verification determines whether $\sigma$ is an element of $L(\varphi)$, where $L(\varphi)$ denotes the set of traces that satisfy the property $\varphi$.
From this perspective, runtime verification corresponds to a word problem: checking whether a given word belongs to a language.
Since these tools run alongside the target system, performance is a primary requirement, and they need lightweight monitoring techniques.

The properties monitored by RV tools are expressed using formal specification languages. 
These languages differ in expressiveness, monitorability, and computational efficiency.
Most of these languages are based on temporal logic: \textit{Linear Temporal Logic (LTL)} expresses time-dependent properties by combining logical propositions with temporal operators~\cite{pnueli1977temporal}, \textit{Metric Temporal Logic (MTL)} extends LTL with quantitative timing constraints for real-time requirements~\cite{koymans1990specifying}, and \textit{Metric First-Order Temporal Logic (MFOTL)} extends MTL further to cover data-carrying events~\cite{basin2015monitoring}.

RV tools vary in the specification formalisms they support, the monitoring algorithms they use, and the internal representations used to evaluate specifications.
MonPoly, for instance, is an RV tool that checks events from data-rich timestamped logs against properties written in the MFOTL specification language~\cite{basin2017monpoly,basin2015monitoring}.
It can be used both offline (reading events from a log file) and online (processing standard input), and it is implemented in OCaml.
MonPoly uses finite relations and relational algebra to perform monitoring.
Its monitoring procedure is based on a bottom-up evaluation of the formula: the monitor first computes the results of subformulas and then reuses them when evaluating larger components of the formula.

Reelay is another RV tool that supports multiple specification formalisms, including past-time LTL, MTL, STL, and first-order quantification~\cite{ulus2026online,ulus2026reelay}.
It provides both online and offline monitoring, as well as dense-time and discrete-time semantics, with a focus on efficiency.
It is implemented as a header-only C++ library and offers its own specification language for defining properties.
Reelay uses sequential networks as the internal representation of formulas, enabling optimizations over them.
Similar to DejaVu~\cite{havelund2018dejavu}, another RV tool that provides a BDD-based monitor for first-order past-time properties, Reelay evaluates properties incrementally via expansion laws relating each subformula's current and previous valuations.

RV tools also differ in the event data formats they consume.
Reelay, for example, uses JSON to represent monitored events.
Other RV tools may rely on their own textual trace formats or structured message representations.
In performance-sensitive settings, the choice of event representation affects both the amount of data transmitted and the processing required before monitoring.
Text-based formats such as JSON and XML are widely used for their readability, but binary formats can provide more compact representations and reduce serialization overhead.
Two commonly used binary formats, both developed by Google, are \textit{Protocol Buffers}~\cite{protobuf} and \textit{FlatBuffers}~\cite{flatbuffers}.
Protocol Buffers encodes fields compactly and parses them into generated message objects before individual fields can be read, making it well-suited for network transmission.
FlatBuffers, in contrast, enables zero-copy access to fields directly within the serialized buffer via offset tables, favoring low-latency access at the cost of a more involved construction step on the writing side.

RV tools further differ in the mechanisms through which they integrate with the autonomous systems they monitor.
Autonomous systems are commonly organized as distributed components that exchange data while interacting with their environment.
Pub/sub networks provide a flexible communication model for this purpose, decoupling publishers from subscribers.
A well-known example is the Robot Operating System (ROS)~\cite{quigley2009ros}, in which components are organized as nodes that communicate over a pub/sub infrastructure.
Such systems can be monitored at runtime by observing the event streams exchanged between these components.
RV tools such as ROSRV~\cite{huang2014rosrv} and ROSMonitoring~\cite{ferrando2020rosmonitoring} follow this approach for ROS-based autonomous systems, collecting monitored events through the pub/sub infrastructure.
The communication mechanism, therefore, becomes part of the monitoring setup, as the RV tool receives the event stream through the same infrastructure used by the monitored system.
This pub/sub mechanism is provided by IPC libraries; two commonly used ones are \textit{iceoryx2}~\cite{iceoryx2} and \textit{Zenoh}~\cite{zenoh}.
iceoryx2 is a community-driven IPC library designed for communication between processes on the same machine over shared memory.
It avoids data copies and locks in the transport path, making it suitable for systems with low-latency requirements.
Zenoh is a communication middleware that supports pub/sub messaging both over shared memory on the same host and over the network across distributed hosts.
It is well-suited for multi-host and containerized architectures, as well as for testing network-based communication on a single machine.

A wide variety of RV tools exists, and the community has recognized the need to classify and evaluate them, through a taxonomy of RV tools~\cite{falcone2021taxonomy} and efforts such as the Runtime Verification Benchmark Challenge~\cite{rvbenchmarkchallenge2018}.
For RV tools operating over pub/sub infrastructures, evaluation must consider not only their correctness and expressive power for specifying and monitoring a property, but also how efficiently and reliably they process events delivered through the communication infrastructure.
In such settings, performance depends on both the monitoring procedure and the surrounding event-processing pipeline, including the representation and transport of monitored events.
This makes systematic evaluation challenging, since isolated performance numbers are difficult to interpret or compare across tools and communication settings.
Benchmarking addresses this by providing a controlled, repeatable basis for comparison, defining common specifications, traces, event representations, communication mechanisms, and measurement criteria under which different tools can be evaluated.

This paper presents RyBench, a benchmarking framework for RV tools over pub/sub networks that acts as a synthetic event publisher in place of a monitored system.
The main contributions of this work are:
\begin{itemize}
\item A modular benchmarking framework separating event generation, serialization, and transport, providing a set of implementations that can be combined and extended independently.
\item Online and offline execution modes, including a fixed generator that replays recorded traces for reproducible testing and can convert them across serialization formats.
\item An experimental evaluation measuring RyBench's throughput using Open Simulation Interface (OSI) schemas and Timescales properties, and demonstrating the end-to-end workflow through an application to Reelay~\cite{ulus2026reelay} and an integration with MonPoly.
\end{itemize}

The remainder of the paper is organized as follows.
Section~\ref{sec:related-work} surveys related work on RV tool evaluation, benchmark generation, and RV over pub/sub networks.
Section~\ref{sec:architecture} describes RyBench's architecture and its components.
Section~\ref{sec:evaluation} reports RyBench's throughput performance across its components and evaluates Reelay using RyBench, while also demonstrating its integration with MonPoly.
Section~\ref{sec:conclusion} discusses the findings of this work and points to possible directions for future work.








\section{Related Work}
\label{sec:related-work}

Several efforts have classified and evaluated RV tools.
A taxonomy~\cite{falcone2021taxonomy} classifies RV tools by specification, deployment, and monitoring aspects, and an earlier community effort collected requirements and recommendations for evaluating them~\cite{reger2017report,bartocci2019crv}.
RyBench shares the motivation behind these efforts: the systematic evaluation of RV tools.

Timescales~\cite{ulus2019timescales} generates traces for ten selected MTL properties with parametric time bounds, each given in both a past-time and a future-time variant.
Its generation time is largely invariant to the time bounds and therefore remains efficient even for large bounds; it is implemented in Python and generates about 1 million events per second.
We adopt three of the Timescales properties to check RyBench's traces with Reelay.
In contrast to Timescales, which generates traces offline in CSV format, RyBench provides end-to-end benchmarking with multiple data formats.

Another benchmark generator~\cite{krstic2020benchmark} targets specifications written in MFODL~\cite{basin2020formally}.
Their tool generates both satisfying and violating traces for a multi-way join formula, while the oracle provides the expected verdicts: a stream generator produces random event traces, and a replayer sends previously generated traces to an online monitor at a configurable rate.
RyBench's schema-guided random generator is similar to their random mode, but it supports user-defined schemas with arbitrary attributes; likewise, our fixed generator follows the same general idea as their replayer, and we additionally use it to test RV tools over past-time temporal logic and to generate high event loads.

A different approach~\cite{li2018mltl} generates benchmarks via \emph{formula progression}, a technique originating from planning~\cite{bacchus1998planning}, producing finite traces for MLTL~\cite{reinbacher2014temporal} formulas together with an oracle that records the truth value of the formula at each step.
Their algorithm computes progressed formulas and invokes an MLTL-SAT solver~\cite{li2019satisfiability} to obtain trace extensions, which are appended to the actual trace, and provides three generation patterns that control the distribution of false verdicts within the oracle.
The follow-up tool FPROGG~\cite{rosentrater2025fprogg} implements this formula-progression approach as an open-source C++ generator.
Their solver-based approach, however, invokes a full decision procedure at every step, which is well-suited to offline generation of complex formulas but is too costly for the online, lightweight generation RyBench requires.

Trace generation for temporal properties is closely related to both the long-studied classical satisfiability and sampling problems.
Bounded LTL and MTL satisfiability can be encoded into the Boolean satisfiability problem~\cite{cook1971complexity,levin1973universal}, decidable via SAT and SMT solvers~\cite{davis1962machine,demoura2008z3,barbosa2022cvc5}, by tableau construction~\cite{wolper1985tableau}, or by reduction to model checking~\cite{rozier2010ltl}; MTL and MLTL satisfiability have been studied extensively and further in~\cite{ouaknine2005decidability,alur1996benefits,li2019satisfiability}.
Beyond deciding existence, sampling a satisfying trace with guarantees requires estimating or partitioning the solution space~\cite{valiant1979complexity,chakraborty2013scalable,barbot2016uniform,barbot2023wordgen}, and recent work extends this line to signal temporal logic through a satisfiability tableau that yields witnesses~\cite{melani2025tree}, and to timed regular expressions through volume-based uniform sampling~\cite{gigler2025uniform}.
These approaches provide completeness or distributional guarantees at the cost of a decision procedure per step; high-throughput benchmark generation instead requires trading these guarantees for speed at scale, in the same spirit as stochastic local search solvers~\cite{selman1992gsat,selman1994walksat}.

Beyond generation, RV performance has been studied from several angles.
A synthetic benchmarking framework generates concurrent master-slave monitoring scenarios and message bursts, reporting system-level utilization metrics alongside RV-specific ones~\cite{aceto2021benchmarking}.
An in-depth study of RV overhead during software testing finds that instrumentation, rather than monitoring itself, dominates the overhead in JavaMOP-based unit tests~\cite{guan2024depth,jin2012javamop}, and RV tool efficiency has separately been evaluated at scale over real-world automated test suites~\cite{javed2018large}.
In real-time settings, event- and time-triggered monitoring has been used to reduce and bound runtime overhead~\cite{wu2013reducing}.

RV over pub/sub networks is the subject of several earlier works.
ROSRV~\cite{huang2014rosrv} and ROSMonitoring~\cite{ferrando2020rosmonitoring,saadat2024rosmonitoring} integrate RV monitors directly into the existing pub/sub messaging infrastructure of ROS-based systems, with ROSMonitoring reporting a monitoring overhead of a few multiples for message streams in the low kHz range.
More recently, Runtime Verification Containers instead target the pub/sub layer itself, deploying containerized monitors over DDS and Zenoh and measuring the resulting messaging throughput and latency~\cite{mehran2024runtime}, whereas RyBench generates the benchmark workload itself, with controllable generators, serializers, and transports.

Beyond the RV context, other studies compare the performance of serialization formats~\cite{viotti2022benchmark,maltsev2024impact,jackson2024streaming} and pub/sub middlewares~\cite{liang2023performance}, showing that both are decisive factors for messaging and parsing performance, while shared-memory transport designs such as iceoryx target the same overheads by construction~\cite{pohnl2023shared}.
Closer to RyBench's own transport layer, synthetic pub/sub load generators such as Apex.AI's \texttt{performance\_test}~\cite{apexai2019performancetest} and iRobot's \texttt{ros2-performance}~\cite{irobot2019ros2performance} offer configurable message rates and sizes and latency statistics reporting across ROS~2 middlewares.
While those works provide benchmarking methodologies and results for their own purposes, RyBench provides a common benchmarking infrastructure for measuring these factors specifically for RV over pub/sub.

\section{RyBench Framework}
\label{sec:architecture}
 
\subsection{Architecture}
\label{sec:arch-overview}
 
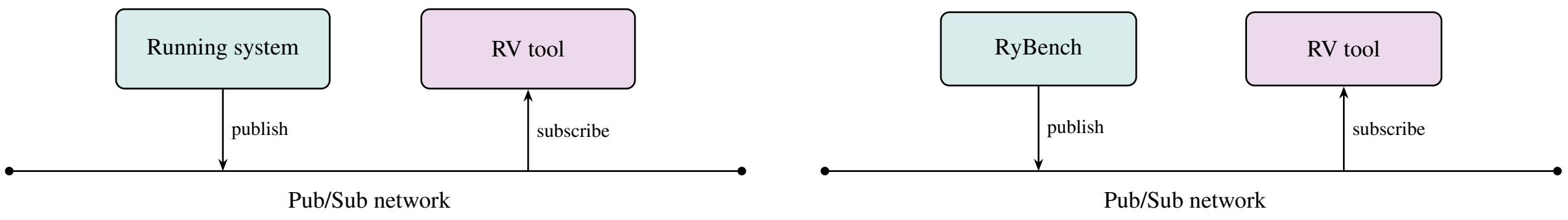
\begin{figure}[ht]
    \centering
    \begin{minipage}[t]{0.48\textwidth}
        \centering
        \resizebox{\linewidth}{!}{
        \begin{tikzpicture}[
            box/.style={
                draw, rounded corners, minimum width=3.5cm, minimum height=1.3cm,
                align=center, thick
            },
            pub/.style={box, fill=teal!15},
            sub/.style={box, fill=violet!15},
            arr/.style={-{Stealth[length=2mm]}, thick}
        ]
            \draw[thick] (-0.5,0) -- (11.5,0);
            \fill (-0.5,0) circle (2pt);
            \fill (11.5,0) circle (2pt);
            \node[below] at (5.4,-0.2) {Pub/Sub network};
            \node[pub] (sys) at (3,2) {Running system};
            \draw[arr] (sys.south) -- node[right]{\footnotesize publish} (3,0);
            \node[sub] (rv) at (8,2) {RV tool};
            \draw[arr] (8,0) -- node[right]{\footnotesize subscribe} (rv.south);
        \end{tikzpicture}
        }
        \subcaption{RV tool over a pub/sub network.}
        \label{fig:pubsub-standard}
    \end{minipage}
    \hfill
    \begin{minipage}[t]{0.48\textwidth}
        \centering
        \resizebox{\linewidth}{!}{
        \begin{tikzpicture}[
            box/.style={
                draw, rounded corners, minimum width=3.2cm, minimum height=1.2cm,
                align=center, thick
            },
            pub/.style={box, fill=teal!15},
            sub/.style={box, fill=violet!15},
            arr/.style={-{Stealth[length=2mm]}, thick}
        ]
            \draw[thick] (-0.5,0) -- (11.5,0);
            \fill (-0.5,0) circle (2pt);
            \fill (11.5,0) circle (2pt);
            \node[below] at (5.4,-0.2) {Pub/Sub network};
            \node[pub] (sys) at (3,2) {RyBench};
            \draw[arr] (sys.south) -- node[right]{\footnotesize publish} (3,0);
            \node[sub] (rv) at (8,2) {RV tool};
            \draw[arr] (8,0) -- node[right]{\footnotesize subscribe} (rv.south);
        \end{tikzpicture}
        }
        \subcaption{RyBench over a pub/sub network.}
        \label{fig:pubsub-rybench}
    \end{minipage}
    \caption{Pub/sub architecture for monitoring and benchmarking.}
    \label{fig:pubsub-network}
\end{figure}

A pub/sub network, in a standard setup, serves as the primary communication channel between the running system and the RV tool, with event messages flowing from the system to the tool (Figure~\ref{fig:pubsub-standard}).
The RV tool monitors the system through these event messages and has no other access to its internal state, so the event messages are the only available source of information about the running system.
RyBench mocks the running system by generating and publishing such event messages itself; the RV tool then receives them from RyBench over the same pub/sub network, rather than from the actual system (Figure~\ref{fig:pubsub-rybench}).

\begin{figure}[ht]
\centering
\begin{tikzpicture}[
  font=\small,
  node distance=1cm,
  box/.style={draw, rounded corners, align=center,
  minimum height=0.9cm, fill=blue!4},
  widebox/.style={box, minimum width=3.2cm, minimum height=1.7cm},
  arrow/.style={-{Latex[length=2.5mm]}, thick},
  publisherbox/.style={draw, dashed, rounded corners=10pt, fill=gray!6,
    inner xsep=12pt, inner ysep=10pt}
]
\node[widebox] (generator) {\\\\Event\\Generator\\\\};
\node[widebox, right=1.5cm of generator] (serializer) {\\\\Message\\Serializer\\\\};
\node[widebox, right=1.5cm of serializer] (transport) {\\\\Transport\\Layer\\\\};
\draw[arrow] (generator.east) -- (serializer.west);
\draw[arrow] (serializer) -- (transport);
\coordinate (publisher-bottom) at ([yshift=-0.4cm]transport.south);
\coordinate (publisher-up) at ([yshift=0.4cm]transport.north);
\begin{pgfonlayer}{background}
\node[publisherbox, fit=(generator)(serializer)(transport)(publisher-bottom)(publisher-up)] (publisher) {};
\end{pgfonlayer}
\end{tikzpicture}
\caption{RyBench components.}
\label{fig:publisher-node-pipeline}
\end{figure}
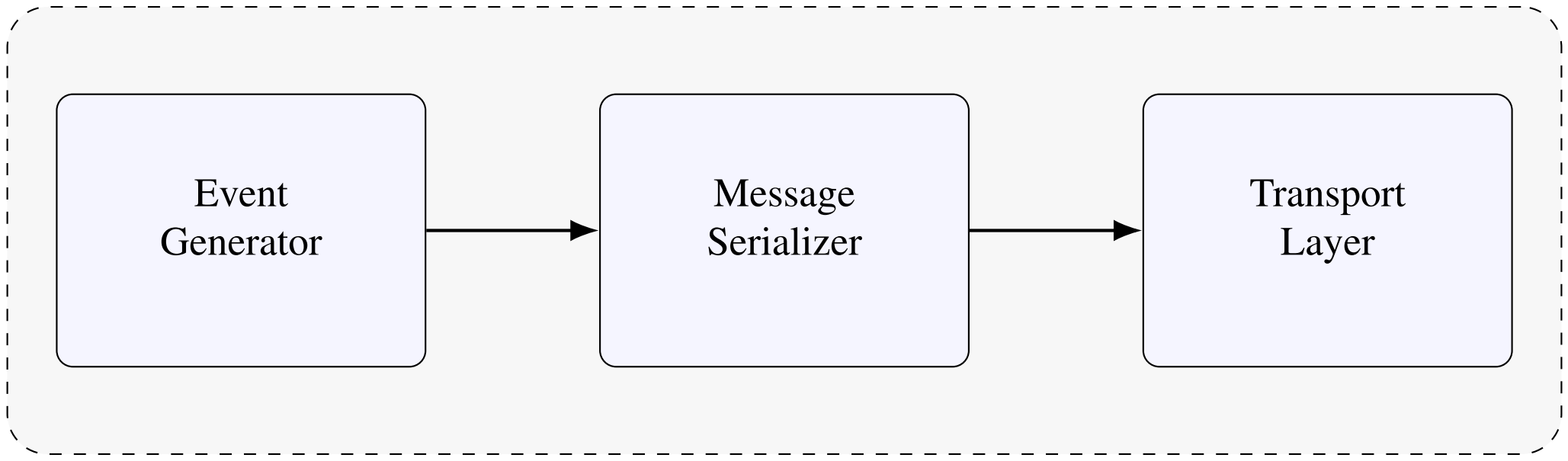

RyBench divides the task of generating and publishing events into three sub-tasks, each handled by a separate component: an event generator produces events according to a scheme that defines the benchmark scenario; a message serializer converts events into a format the RV tool can deserialize; and a transport layer publishes the serialized messages over the pub/sub network (Figure~\ref{fig:publisher-node-pipeline}).
This modular structure makes RyBench adaptable to different scenarios, formats, and transports, since one implementation of a component can be substituted for another without structural change.
 
RyBench is implemented in C++23, whose compiled, native execution and direct memory control enable benchmark generation at high event rates and low latency.
The generated values are appended into the buffer via the serializer component without an intermediate copy, and for zero-copy transports the message is written directly into a buffer the transport provides.
Each component is defined as an abstract interface, so that new implementations can be added without changing the other components or the overall pipeline.

\subsection{Event Generation}
\label{sec:arch-generation}
 
RyBench implements two generation approaches: a random generator for basic functional tests and a fixed generator for reproducing existing benchmarks.
A third approach, property-guided generation, is proposed as an extension to generate events according to a given specification.
 
The first technique is random generation, where the generator samples values at each time step from a distribution over given limits, independent of previously generated values.
The generated events are syntactically valid but carry no semantic or temporal meaning, so this generator is suitable only for testing primary functionalities of an RV tool, such as message parsing, connection handling, and behavior under high load.
It receives field names, types, and field-specific attributes, such as minimum and maximum bounds and roles, directly from the serialization component.
It supports both primitive and composite types.
For each field, it samples a value and passes it directly to the serialization backend, which builds the message.
No intermediate shared data structure is used to hold a generated event, avoiding an extra copy and keeping the data path efficient.
 
The second technique is fixed generation, where event traces are reproduced identically by reading previously recorded messages.
It is the basis for regression testing: a version-controlled fixture can be fed to an RV tool across successive versions or configurations, and any deviation in the tool's verdicts, timing, or outputs is attributable to the system under test rather than to a change in the input.
The same property enables A/B testing, where one trace is replayed across different transports or serializers and the results are compared.
Internally, the generator reads a binary dump of length-prefixed records, each tagged with its wire format; the serialization layer forwards these bytes directly whenever the tag matches the target, keeping the path minimal for high-load testing, and takes responsibility for the conversion only when it is required.

The third technique is property-guided generation, where event traces are supposed to be generated under the guidance of a temporal logic property.
It needs to obtain variables and their related specifications from the schema and generate events one by one using a trace generation algorithm.
Then it needs to pass them to the serialization component to create the actual event message.
Existing benchmark generators use various approaches for generating traces offline.
Timescales~\cite{ulus2019timescales} provides an implementation for ten selected MTL properties with parametric bounds.
The generator of Krstić and Schneider~\cite{krstic2020benchmark} provides a formula-specific mode for a multi-way join formula in MFODL.
FPROGG~\cite{rosentrater2025fprogg}, following Li and Rozier's~\cite{li2018mltl} formula-progression approach, supports arbitrary MLTL formulas by using an MLTL-SAT solver.
More generally, since SAT-based~\cite{davis1962machine} and tableau-based~\cite{wolper1985tableau} methods also produce traces as satisfying witnesses, their algorithms could be adopted as alternative generator implementations within RyBench's modular architecture for online generation.

\subsection{Message Serialization}
\label{sec:arch-serialization}

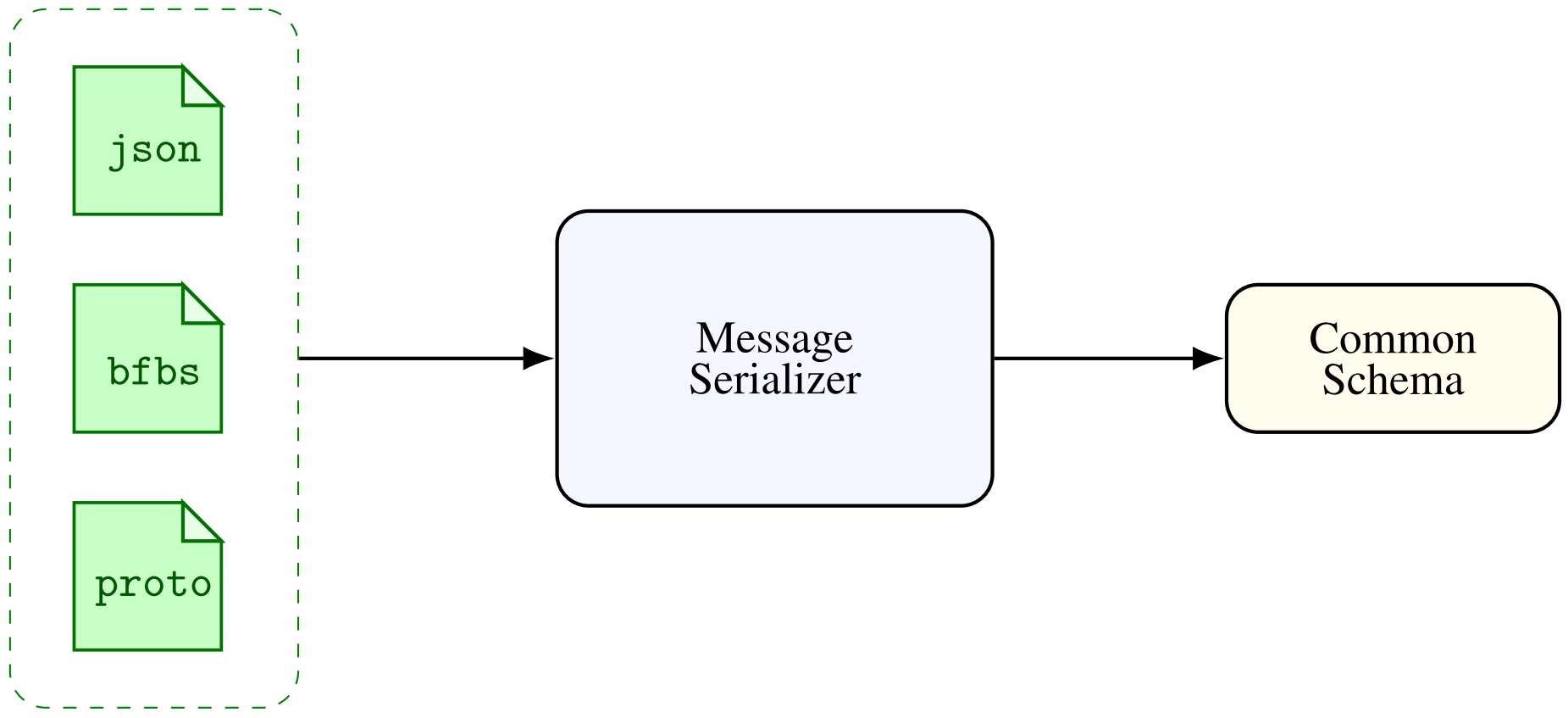
\begin{figure}[!b]
\centering
\begin{tikzpicture}[
  font=\small,
  >=Latex,
  component/.style={
    draw, rounded corners=7pt,
    minimum width=3.4cm, minimum height=2.3cm,
    align=center, fill=blue!4, thick
  },
  common/.style={
    draw, rounded corners=7pt,
    minimum width=2.6cm, minimum height=1.15cm,
    align=center, fill=yellow!8, thick
  },
  group/.style={
    draw, dashed, rounded corners=8pt,
    inner xsep=0.3cm, inner ysep=0.40cm,
    draw=green!50!black
  },
  arrow/.style={-{Latex[length=2.5mm]}, thick}
]
\tikzset{
  file/.pic={
    \path[
      pic actions,
      draw=green!45!black,
      line width=0.7pt,
      fill=green!22
    ]
      (0,0) -- (0,1.15) -- (0.85,1.15) -- (1.15,0.85) -- (1.15,0) -- cycle;
    \draw[
      green!45!black,
      line width=0.8pt,
      fill=green!10
    ]
      (0.85,1.15) -- (0.85,0.85) -- (1.15,0.85) -- cycle;
    \node[
      font=\small\bfseries,
      text=green!35!black
    ] at (0.625,0.48) {#1};
  }
}
\pic (json)  at (0, 1.8) {file={\texttt{json}}};
\pic (bfbs)  at (0, 0.1) {file={\texttt{bfbs}}};
\pic (proto) at (0,-1.6) {file={\texttt{proto}}};
\coordinate (schemas-top) at (-0.2,3);
\coordinate (schemas-bottom) at (1.45,-1.65);
\begin{pgfonlayer}{background}
  \node[group, fit=(schemas-top)(schemas-bottom)] (schemas) {};
\end{pgfonlayer}
\node[component, right=2cm of schemas] (serializer)
  {Message\\[-0.1cm]Serializer};
\node[common, right=1.8cm of serializer] (common)
  {Common\\[-0.1cm]Schema};
\draw[arrow] (schemas.east) -- (serializer.west);
\draw[arrow] (serializer.east) -- (common.west);
\end{tikzpicture}
\caption{Schema conversion.}
\label{fig:serialization-schema}
\end{figure}

The serialization component converts events into messages in a target wire format, and is responsible for message construction during generation and format conversion during replay.
It parses a format-specific schema, such as a \texttt{.json} file, a compiled FlatBuffers \texttt{.bfbs} file, or a Protobuf \texttt{.proto} file, into a common schema, a format-independent representation of the message fields, types, and attributes such as value ranges or guiding properties (Figure~\ref{fig:serialization-schema}).
The generator relies on this common schema when producing events, with no intermediate message representation beyond it, so it remains agnostic to the underlying wire format. 
Each message is built in its target format without extra copy, keeping the serialization path lightweight.

The common usage of FlatBuffers and Protocol Buffers relies on code generation at compile time, where a schema compiler produces format-specific classes for a fixed message type.
Since the framework must work with different schemas at runtime rather than a single, fixed one, it instead loads schemas through a reflection interface: FlatBuffers through its reflection API, and Protocol Buffers through its dynamic message API.
As a result, a new schema can be introduced without generating or compiling format-specific code beforehand.

RyBench implements three message formats, chosen for three different design philosophies: JSON, a text-based format already used by RV tools such as Reelay but with non-trivial parsing cost; FlatBuffers, which provides zero-copy access suited to low-latency, same-host monitoring; and Protocol Buffers, a compact binary format suited to network transmission.
Supporting all three lets the framework measure how the choice of format affects the end-to-end performance of an RV tool.

The serializer exposes a field-by-field value append interface across all three formats.
For JSON and Protocol Buffers, each value is appended directly to the message being built; FlatBuffers, in contrast, requires bottom-up construction, since a table can only be finalized after every object it references already exists in the buffer, so the FlatBuffers serializer holds values on a per-level stack when appending and emits them in reverse after all values are available.
Despite this internal difference, the append interface itself remains uniform across formats, so the generator does not need to account for it.

Beyond message construction, the component can also convert a message from one format to another at runtime, which lets the same recorded trace be replayed across different formats.
The conversion is not equally straightforward for every format pair, since their type systems do not always align.
Between FlatBuffers and Protocol Buffers, \texttt{ubyte} and \texttt{ushort} have no direct Protocol Buffers counterparts and must be widened. FlatBuffers structs and unions do not correspond to Protocol Buffers messages and \texttt{oneof} fields. 
Hence, the mapping is approximate and guarantees related to inline layout or explicit union tags are not preserved.
Conversions involving JSON, by contrast, are handled through existing library support.
 
\subsection{Transport Layer}
\label{sec:arch-transport}

The transport layer transmits serialized event messages to the RV tool over a pub/sub channel.
By abstracting the underlying transmission mechanism, this approach enables different implementations to be used in a pluggable manner, making it compatible with RV tools that use different transmission layers.
Because it is decoupled from event generation and serialization, an identical stream of events can be produced regardless of how it is transmitted, which lets the transmission overhead itself be measured in isolation, and enables end-to-end benchmarking that covers message consumption, parsing, and verdict computation together.

\begin{figure}[!b]
    \centering
    \begin{subfigure}[b]{0.49\textwidth}
        \centering
        \resizebox{\linewidth}{!}{%
        \begin{tikzpicture}[scale=1.25, every node/.style={transform shape},
            box/.style={draw, rounded corners, minimum width=3.4cm, minimum height=1.3cm, align=center, thick},
            pub/.style={box, fill=teal!15},
            sub/.style={box, fill=violet!15},
            arr/.style={-{Stealth[length=2mm]}, thick}
        ]
            \useasboundingbox (-0.5,-0.6) rectangle (11.5,2.9);
            \draw[thick] (-0.5,0) -- (11.5,0);
            \fill (-0.5,0) circle (2pt); \fill (11.5,0) circle (2pt);
            \node[below] at (5.5,-0.2) {Pub/Sub network};
            \node[pub] (sys) at (3,2) {RyBench};
            \node[sub] (rv) at (8,2) {RV tool};
            \draw[arr] (sys.south) -- node[right]{\footnotesize publish} (3,0);
            \draw[arr] (8,0) -- node[right]{\footnotesize subscribe} (rv.south);
        \end{tikzpicture}%
        }
        \subcaption{Online generation -- same host.}
        \label{fig:gen-online-same}
    \end{subfigure}
    \hfill
    \begin{subfigure}[b]{0.49\textwidth}
        \centering
        \resizebox{\linewidth}{!}{%
        \begin{tikzpicture}[scale=1.25, every node/.style={transform shape},
            box/.style={draw, rounded corners, minimum width=3.4cm, minimum height=1.3cm, align=center, thick},
            pub/.style={box, fill=teal!15},
            sub/.style={box, fill=violet!15},
            arr/.style={-{Stealth[length=2mm]}, thick},
            host/.style={draw, dashed, rounded corners, inner ysep=17pt, inner xsep=14pt}
        ]
            \useasboundingbox (-0.5,-0.6) rectangle (11.5,2.9);
            \node[pub] (sys) at (3,2) {RyBench};
            \node[sub] (rv) at (8,2) {RV tool};
            \node[host, fit=(sys),yshift=0.5mm] (hA) {};
            \node[host, fit=(rv),yshift=0.5mm]  (hB) {};
            \node[font=\footnotesize] at ([yshift=-6pt,xshift=-42pt]hA.north) {host A};
            \node[font=\footnotesize] at ([yshift=-6pt,xshift=-42pt]hB.north) {host B};
            \draw[thick] (-0.5,0) -- (11.5,0);
            \fill (-0.5,0) circle (2pt); \fill (11.5,0) circle (2pt);
            \node[below] at (5.5,-0.2) {Pub/Sub network};
            \draw[arr] (sys.south) -- node[right]{\footnotesize publish} (3,0);
            \draw[arr] (8,0) -- node[right]{\footnotesize subscribe} (rv.south);
        \end{tikzpicture}%
        }
        \subcaption{Online generation -- different host.}
        \label{fig:gen-online-diff}
    \end{subfigure}
    \vspace{0.5em}
    \begin{subfigure}[b]{0.49\textwidth}
        \centering
        \vspace{1.2em}
        \resizebox{\linewidth}{!}{%
        \begin{tikzpicture}[scale=1.25, every node/.style={transform shape},
            box/.style={draw, rounded corners, minimum width=3.4cm, minimum height=1.3cm, align=center, thick},
            pub/.style={box, fill=teal!15},
            sub/.style={box, fill=violet!15},
            arr/.style={-{Stealth[length=2mm]}, thick},
            pics/db/.style={code={
            \def\dbr{0.55}
            \def\dbh{1.0}
            \def\dbe{0.18}
            \path[draw=gray!45!black, line width=0.9pt, fill=gray!18]
              (-\dbr,0) -- (-\dbr,\dbh)
              arc[start angle=180, end angle=360, x radius=\dbr, y radius=\dbe]
              -- (\dbr,0)
              arc[start angle=0, end angle=-180, x radius=\dbr, y radius=\dbe];
            \path[draw=gray!45!black, line width=0.9pt, fill=gray!28]
              (0,\dbh) ellipse[x radius=\dbr, y radius=\dbe];
            \draw[gray!45!black, line width=0.6pt]
              (-\dbr,\dbh*0.66) arc[start angle=180, end angle=360, x radius=\dbr, y radius=\dbe];
            \draw[gray!45!black, line width=0.6pt]
              (-\dbr,\dbh*0.33) arc[start angle=180, end angle=360, x radius=\dbr, y radius=\dbe];
            }}
        ]
        \useasboundingbox (-0.5,-2.0) rectangle (11.5,0.7);
        \node[pub] (sys) at (1.7,0) {RyBench};
        \pic (tf) at (5.45,-1.5) {db};
        \node[sub] (rv) at (9.3,0) {RV tool};
        \draw[arr] (sys.east) to[bend left=10] node[above]{\footnotesize write} (5.35,-0.45);
        \draw[arr] (5.55,-0.45) to[bend left=10] node[above]{\footnotesize read} (rv.west);
        \end{tikzpicture}%
        }
        \subcaption{Offline generation}
        \label{fig:gen-offline}
    \end{subfigure}
    \caption{RyBench generation modes: online (same/different host) and offline.}
    \label{fig:generation-modes}
\end{figure}
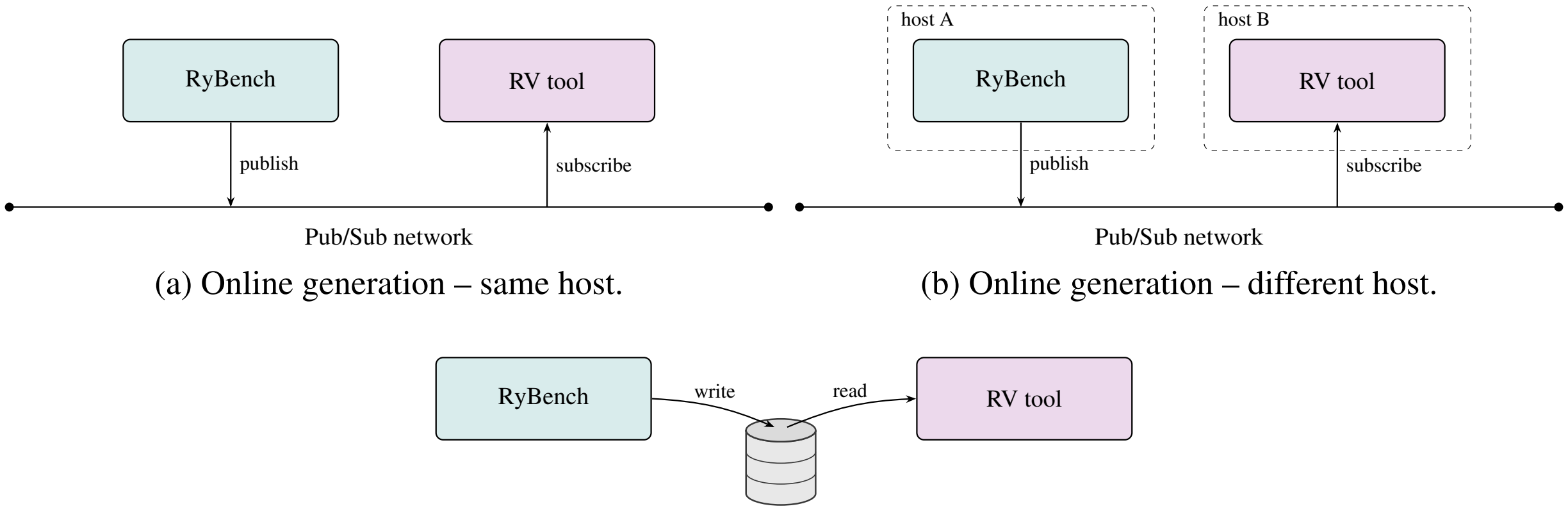

The choice of transport determines whether RyBench operates as an online or an offline benchmark generator (Figure~\ref{fig:generation-modes}).
In its online mode, RyBench publishes messages immediately over the pub/sub network as they are generated, either on the same host or across separate hosts, depending on configuration.
In its offline mode, the generated messages are persisted to a file instead; the resulting file can then be replayed by the fixed generator (Section~\ref{sec:arch-generation}) to reproduce the same trace, or fed directly into the offline modes of an RV tool.

RyBench provides two pub/sub transport implementations for online generation: Zenoh and iceoryx2.
Zenoh mainly targets multi-host systems but also supports multiple transport mechanisms and can be used on a single host to make the cost of networking explicit.
iceoryx2 uses shared memory for zero-copy transfer between processes on the same host, with configurable buffering and delivery policies.
For offline generation, filewriter writes the event messages to a file instead of publishing them.

\begin{figure}[!t]
    \centering
    \begin{subfigure}[t]{0.48\textwidth}
        \centering
        \resizebox{\linewidth}{!}{
        \begin{tikzpicture}[scale=1.25, every node/.style={transform shape},
            box/.style={draw, rounded corners, minimum width=3.4cm, minimum height=1.3cm, align=center, thick},
            pub/.style={box, fill=teal!15},
            sub/.style={box, fill=violet!15},
            arr/.style={-{Stealth[length=2mm]}, thick}
        ]
            \draw[thick] (-0.5,0) -- (11.5,0);
            \fill (-0.5,0) circle (2pt); \fill (11.5,0) circle (2pt);
            \node[below] at (5.5,-0.2) {Pub/Sub network};
            \node[pub] (p) at (3,2) {RyBench Node};
            \node[sub] (s) at (8,2) {RV tool Node};
            \draw[arr] (p.south) -- node[right]{\footnotesize publish} (3,0);
            \draw[arr] (8,0) -- node[right]{\footnotesize subscribe} (s.south);
        \end{tikzpicture}}
        \subcaption{One-to-one.}
        \label{fig:topo-one-to-one}
    \end{subfigure}
    \hfill
    \begin{subfigure}[t]{0.48\textwidth}
        \centering
        \resizebox{\linewidth}{!}{
        \begin{tikzpicture}[scale=1.25, every node/.style={transform shape},
            box/.style={draw, rounded corners, minimum width=3.4cm, minimum height=1.3cm, align=center, thick},
            pub/.style={box, fill=teal!15},
            arr/.style={-{Stealth[length=2mm]}, thick},
            pics/stacksub/.style={code={
                \draw[rounded corners, thick, fill=violet!15] (-1.54,-0.49) rectangle (1.86,0.81);
                \draw[rounded corners, thick, fill=violet!15] (-1.62,-0.57) rectangle (1.78,0.73);
                \draw[rounded corners, thick, fill=violet!15] (-1.7,-0.65) rectangle (1.7,0.65);
                \node at (0,0) {#1};
            }}
        ]
            \draw[thick] (-0.5,0) -- (11.5,0);
            \fill (-0.5,0) circle (2pt); \fill (11.5,0) circle (2pt);
            \node[below] at (5.5,-0.2) {Pub/Sub network};
            \node[pub] (p) at (3,2) {RyBench Node};
            \pic (s) at (8,2) {stacksub={RV tool Node}};
            \draw[arr] (p.south) -- node[right]{\footnotesize publish} (3,0);
            \draw[arr] (8,0) -- node[right]{\footnotesize subscribe} (8,1.35);
        \end{tikzpicture}}
        \subcaption{One-to-many.}
        \label{fig:topo-one-to-many}
    \end{subfigure}
    \vspace{0.5em}
    \begin{subfigure}[t]{0.48\textwidth}
        \centering
        \vspace{1.2em}
        \resizebox{\linewidth}{!}{
        \begin{tikzpicture}[scale=1.25, every node/.style={transform shape},
            box/.style={draw, rounded corners, minimum width=3.4cm, minimum height=1.3cm, align=center, thick},
            sub/.style={box, fill=violet!15},
            arr/.style={-{Stealth[length=2mm]}, thick},
            pics/stackpub/.style={code={
                \draw[rounded corners, thick, fill=teal!15] (-1.54,-0.49) rectangle (1.86,0.81);
                \draw[rounded corners, thick, fill=teal!15] (-1.62,-0.57) rectangle (1.78,0.73);
                \draw[rounded corners, thick, fill=teal!15] (-1.7,-0.65) rectangle (1.7,0.65);
                \node at (0,0) {#1};
            }}
        ]
            \draw[thick] (-0.5,0) -- (11.5,0);
            \fill (-0.5,0) circle (2pt); \fill (11.5,0) circle (2pt);
            \node[below] at (5.5,-0.2) {Pub/Sub network};
            \pic (p) at (3,2) {stackpub={RyBench Node}};
            \node[sub] (s) at (8,2) {RV tool Node};
            \draw[arr] (3,1.35) -- node[right]{\footnotesize publish} (3,0);
            \draw[arr] (8,0) -- node[right]{\footnotesize subscribe} (s.south);
        \end{tikzpicture}}
        \subcaption{Many-to-one.}
        \label{fig:topo-many-to-one}
    \end{subfigure}
    \hfill
    \begin{subfigure}[t]{0.48\textwidth}
        \centering
        \vspace{1.2em}
        \resizebox{\linewidth}{!}{
        \begin{tikzpicture}[scale=1.25, every node/.style={transform shape},
            arr/.style={-{Stealth[length=2mm]}, thick},
            pics/stackpub/.style={code={
                \draw[rounded corners, thick, fill=teal!15] (-1.54,-0.49) rectangle (1.86,0.81);
                \draw[rounded corners, thick, fill=teal!15] (-1.62,-0.57) rectangle (1.78,0.73);
                \draw[rounded corners, thick, fill=teal!15] (-1.7,-0.65) rectangle (1.7,0.65);
                \node at (0,0) {#1};
            }},
            pics/stacksub/.style={code={
                \draw[rounded corners, thick, fill=violet!15] (-1.54,-0.49) rectangle (1.86,0.81);
                \draw[rounded corners, thick, fill=violet!15] (-1.62,-0.57) rectangle (1.78,0.73);
                \draw[rounded corners, thick, fill=violet!15] (-1.7,-0.65) rectangle (1.7,0.65);
                \node at (0,0) {#1};
            }}
        ]
            \draw[thick] (-0.5,0) -- (11.5,0);
            \fill (-0.5,0) circle (2pt); \fill (11.5,0) circle (2pt);
            \node[below] at (5.5,-0.2) {Pub/Sub network};
            \pic (p) at (3,2) {stackpub={RyBench Node}};
            \pic (s) at (8,2) {stacksub={RV tool Node}};
            \draw[arr] (3,1.35) -- node[right]{\footnotesize publish} (3,0);
            \draw[arr] (8,0) -- node[right]{\footnotesize subscribe} (8,1.35);
        \end{tikzpicture}}
        \subcaption{Many-to-many.}
        \label{fig:topo-many-to-many}
    \end{subfigure}
    \caption{Multi-node benchmarking.}
    \label{fig:topologies}
\end{figure}
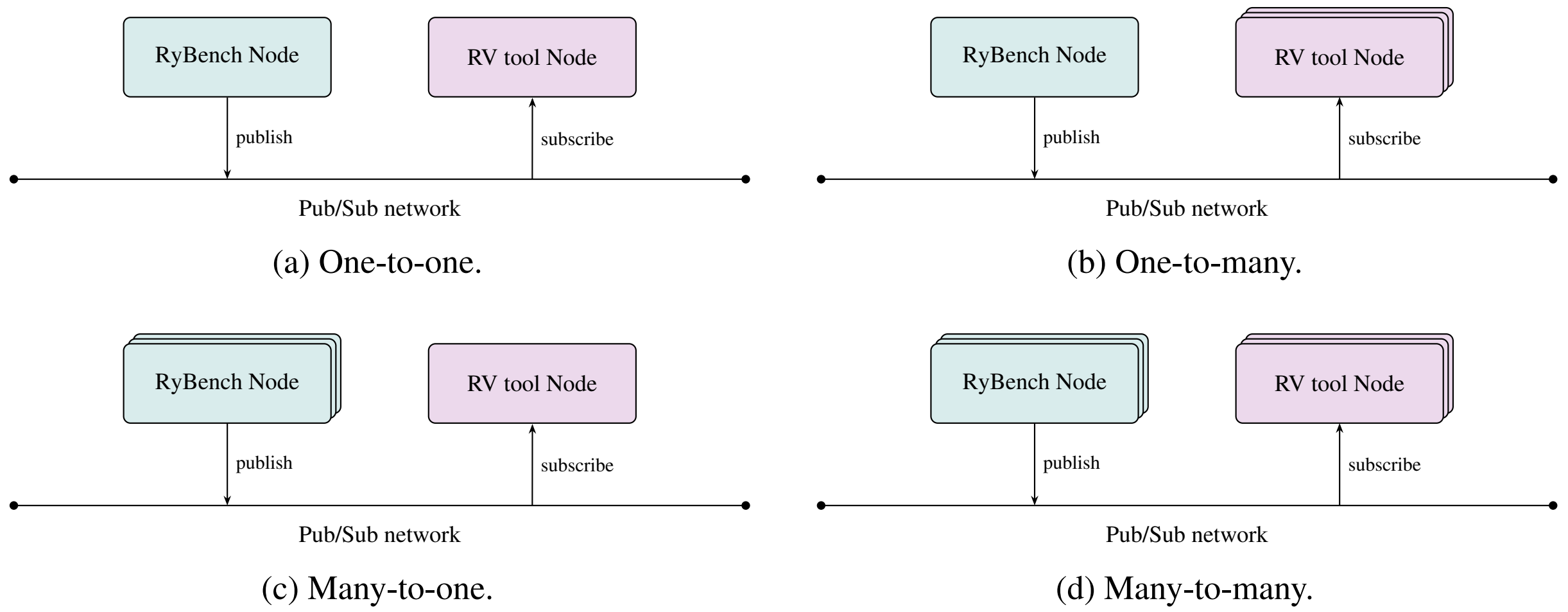

RyBench is also designed to benchmark multi-node monitoring topologies (Figure~\ref{fig:topologies}).
The simplest scenario is one-to-one, where a single RyBench node feeds a single RV tool node.
A single RyBench node can also feed several RV tool nodes simultaneously (one-to-many), with the same event stream observed by multiple monitors.
Conversely, several RyBench nodes can publish into a single RV tool node (many-to-one), mirroring a real system where one monitor watches many components; with multiple RyBench instances, stress tests can vary not only the message rate but also the structure of the publishing nodes.
Finally, several RyBench nodes and RV tool nodes can run simultaneously (many-to-many), thereby reproducing a fully distributed monitoring deployment.

Each node is described by its own configuration -- which transport to use, and settings such as the service or key expression it publishes on, the maximum payload size, and the publishing rate -- from which the framework infers which transport executable to launch and the role the node plays in the topology.
More broadly, RyBench is a configuration-driven framework, in which component implementations are selected both at build time, through compiled-in feature flags, and at run time, through a configuration string.
For multi-node benchmarking, this configuration is stored as a TOML file within a shared configuration directory; its name follows the systemd template convention, falling back from a concrete name to a template whose instance part is substituted into the configuration values as \%i, and a base configuration can be overlaid with node-specific sections that override it.
Several nodes are started at once, by name, through the \texttt{rybench-launch} tool, which resolves each node's configuration, infers its transport, and forks and execs the matching transport executable; every launched node then follows the same lifetime of initialization, a publishing loop, and shutdown.

\section{Evaluation}
\label{sec:evaluation}

This section reports RyBench's experimental evaluation in two parts: its end-to-end publishing throughput capacity across component combinations, to show its capability to be used as an online benchmark generator, and its application to measuring Reelay's monitoring latency and demonstrating the integration of MonPoly.
All experiments were run on a machine with a 13th-generation Intel Core i5-13500H CPU (12 cores -- 4 performance and 8 efficiency cores -- 16 threads) and 16~GB of RAM, running Fedora Linux~42 with Linux kernel~6.19.
The build uses GCC 15.2.1 with \texttt{-std=gnu++23 -O3 -DNDEBUG} in a CMake Release configuration, linking against Reelay v25.0.0, Zenoh 1.7.1, iceoryx2 v0.8.0, FlatBuffers 25.2.10, Protocol Buffers 3.19.6\footnote{Protocol Buffers 3.19.6 is the version
packaged by Fedora~42.}, and RapidJSON (commit \texttt{24b5e7a8}).

\subsection{RyBench}
\label{sec:eval-throughput}

We measured the publishing throughput of RyBench across different combinations of its components: three serializers (JSON, FlatBuffers, Protocol Buffers), two IPC transports (iceoryx2, Zenoh), and two generation algorithms (random generation, fixed generation).
Our aim is to determine whether the framework can sustain the event rates required for realistic testing scenarios.
Throughput is measured at the publisher, covering generation, serialization, and publication, with no subscriber attached, isolating the pipeline's own cost from consumer-side scheduling.
In a separate check, attaching a live subscriber lowered publish throughput by roughly 11--15\% for iceoryx2 and 32\% for Zenoh, without changing the relative orderings reported below for each case.

\begin{figure}[H]
\centering
\begin{minipage}[t]{\textwidth}
\begin{Verbatim}[frame=single,fontsize=\scriptsize,baselinestretch=1.05,label=vehicle\_motion.fbs]
attribute "min";
attribute "max";
table Vector3d {
  x: double (min: "-1000.0", max: "1000.0");
  y: double (min: "-1000.0", max: "1000.0");
  z: double (min: "-100.0",  max: "100.0");
}
table Orientation3d {
  roll:  double (min: "-3.14159", max: "3.14159");
  pitch: double (min: "-3.14159", max: "3.14159");
  yaw:   double (min: "-3.14159", max: "3.14159");
}
table VehicleMotion {
  position:          Vector3d;
  orientation:       Orientation3d;
  velocity:          Vector3d;
  orientation_rate:  Orientation3d;
  acceleration:      Vector3d;
  current_curvature: double (min: "-0.5", max: "0.5");
}
root_type VehicleMotion;
\end{Verbatim}
\end{minipage}
\caption{FlatBuffers schema for \texttt{VehicleMotion}.}
\label{fig:osi-schema-motion}
\end{figure}

\begin{figure}[H]
\centering
\begin{minipage}[t]{\textwidth}
\begin{Verbatim}[frame=single,fontsize=\scriptsize,baselinestretch=1.05,label=osi\_hostvehicledata.fbs]
table HostVehicleData {
  version:              InterfaceVersion;
  timestamp:            Timestamp;
  host_vehicle_id:      Identifier;
  location:             BaseMoving;
  location_rmse:        BaseMoving;
  vehicle_basics:       VehicleBasics;
  vehicle_powertrain:   VehiclePowertrain;
  vehicle_brake_system: VehicleBrakeSystem;
  vehicle_steering:     VehicleSteering;
  vehicle_wheels:       VehicleWheels;
  vehicle_localization: VehicleLocalization;
  vehicle_automated_driving_function: [VehicleAutomatedDrivingFunction];
  vehicle_motion:       VehicleMotion;      // see Figure 7
  route:                Route;
}
root_type HostVehicleData;
\end{Verbatim}
\end{minipage}
\caption{FlatBuffers schema for \texttt{HostVehicleData}.}
\label{fig:osi-schema-hostvehicle}
\end{figure}

For the schema-guided random and fixed generation, we use two message schemas from the Open Simulation Interface (OSI)~\cite{hanke2017generic}, a standard originally developed to describe automated-driving simulation data.
\texttt{HostVehicleData.VehicleMotion} is a flat, six-field message indicating the host vehicle's position, orientation, velocity, orientation rate, acceleration, and curvature.
Although it is a nested message type in the schema hierarchy, it is self-contained and can be serialized on its own.
\texttt{HostVehicleData}, by contrast, is the top-level message.
It is organized into several embedded sub-messages, covering the vehicle's basic parameters, powertrain, brake system, steering, wheels, localization, and motion (among them, \texttt{VehicleMotion} itself), making it representative of large, deeply nested messages.
Figure~\ref{fig:osi-schema-motion} and Figure~\ref{fig:osi-schema-hostvehicle} show their FlatBuffers versions, annotated with the min and max attributes used in RyBench.
OSI defines its messages natively in Protocol Buffers, so that the schema can be used directly, and each serializer loads an equivalent schema in its own format.

\FloatBarrier

Table~\ref{tab:perf-random} reports the end-to-end throughput for random generation, where avg B is the average message size in bytes, msg/s is the throughput in messages per second, and MB/s is the throughput in data volume.
The serializer choice dominates: FlatBuffers and Protocol Buffers reach roughly two to three times the message rate of JSON.
Data volume throughput (MB/s) remains largely comparable between the two message sizes for the same serializer and transport, but message rates differ because of their average message sizes.
Protocol Buffers, however, falls short of this expectation: despite its smaller payload, it is further slowed by its runtime, reflection-based dynamic serialization.

Table~\ref{tab:perf-replay} reports the achieved throughput for the fixed generator, which replays pre-serialized messages directly to the transport, bypassing generation and serialization.
The serialization format itself has no effect here; only the resulting payload size does.
The rate for small messages is nearly independent of the payload format: iceoryx2 sustains roughly 6.4 million messages per second and Zenoh roughly 2.5 million.

\begin{table}[ht]\centering
\caption{Throughput of random generation across serializers and transports.}
\label{tab:perf-random}
\vskip 0.3em
\footnotesize
\begin{tabular}{|l|l|r|r|r|r|r|r|}
\hline
\multirow{2}{*}{\textbf{serializer}} & \multirow{2}{*}{\textbf{transport}} & \multicolumn{3}{c|}{\textbf{small (VehicleMotion)}} & \multicolumn{3}{c|}{\textbf{large (HostVehicleData)}} \\\cline{3-8}
 &  & \textbf{avg B} & \textbf{msg/s} & \textbf{MB/s} & \textbf{avg B} & \textbf{msg/s} & \textbf{MB/s} \\\hline
\multirow{2}{*}{\texttt{json}}        & \texttt{iceoryx2} & 487.2 &  \num{440571} & 214.7 & 5425.6 & \num{39344} & 213.5 \\\cline{2-8}
                                      & \texttt{zenoh}    & 487.2 &  \num{360555} & 175.7 & 5425.6 & \num{35830} & 194.4 \\\hline
\multirow{2}{*}{\texttt{flatbuffers}} & \texttt{iceoryx2} & 252.0 & \num{1305981} & 329.1 & 2876.5 & \num{99028} & 284.8 \\\cline{2-8}
                                      & \texttt{zenoh}    & 252.0 &  \num{821509} & 207.0 & 2876.5 & \num{77747} & 223.6 \\\hline
\multirow{2}{*}{\texttt{protobuf}}    & \texttt{iceoryx2} & 158.0 &  \num{983613} & 155.4 & 1575.7 & \num{84314} & 132.9 \\\cline{2-8}
                                      & \texttt{zenoh}    & 158.0 &  \num{680331} & 107.5 & 1575.7 & \num{70830} & 111.6 \\\hline
\end{tabular}
\end{table}

\begin{table}[ht]\centering
\caption{Throughput of fixed generation across serializers and transports.}
\label{tab:perf-replay}
\vskip 0.3em
\footnotesize
\begin{tabular}{|l|l|r|r|r|r|r|r|}
\hline
\multirow{2}{*}{\textbf{serializer}} & \multirow{2}{*}{\textbf{transport}} & \multicolumn{3}{c|}{\textbf{small (VehicleMotion)}} & \multicolumn{3}{c|}{\textbf{large (HostVehicleData)}} \\\cline{3-8}
 &  & \textbf{avg B} & \textbf{msg/s} & \textbf{MB/s} & \textbf{avg B} & \textbf{msg/s} & \textbf{MB/s} \\\hline
\multirow{2}{*}{\texttt{json}}        & \texttt{iceoryx2} & 483.2 & \num{6359336} & \num{3072.9} & 5439.2 & \num{3045615} & \num{16565.7} \\\cline{2-8}
                                      & \texttt{zenoh}    & 483.2 & \num{2452294} & \num{1185.0} & 5439.2 &  \num{751487} &  \num{4087.5} \\\hline
\multirow{2}{*}{\texttt{flatbuffers}} & \texttt{iceoryx2} & 248.0 & \num{6360561} & \num{1577.4} & 2884.6 & \num{4386981} & \num{12654.5} \\\cline{2-8}
                                      & \texttt{zenoh}    & 248.0 & \num{2452543} &  608.2       & 2884.6 &  \num{752133} &  \num{2169.6} \\\hline
\multirow{2}{*}{\texttt{protobuf}}    & \texttt{iceoryx2} & 154.0 & \num{6354896} &  978.7       & 1566.9 & \num{4380876} &  \num{6864.5} \\\cline{2-8}
                                      & \texttt{zenoh}    & 154.0 & \num{2451983} &  377.6       & 1566.9 &  \num{812242} &  \num{1272.7} \\\hline
\end{tabular}
\end{table}

\FloatBarrier

\subsection{Application}
\label{sec:eval-application}

We evaluate Reelay using RyBench and also demonstrate RyBench's applicability to MonPoly.
Since Reelay is provided as a header-only library, we integrated it directly with RyBench's transport layers.
Tools that consume events through their own textual interfaces, such as MonPoly, instead require an adapter between the transport and the tool.
We apply this approach to connect MonPoly to RyBench's pub/sub network through a subscriber-side adapter.

The latency is measured end to end, from the timestamp taken immediately before an event leaves the transport on the producer side to the moment the verdict becomes available to the consumer, using \texttt{std::chrono::steady\_clock} (\texttt{CLOCK\_MONOTONIC} under glibc) on both ends.
Since all experiments run on a single host, this clock is valid across processes without synchronization.
The producer timestamp itself is carried as an 8-byte raw prefix ahead of the serialized message, outside the serialized payload, so it does not affect the reported message sizes.
Taking a timestamp has a fixed cost that is included in latency results. 
We separately measured this cost via a small side check, finding roughly 1--3\% of the sub-microsecond iceoryx2 medians reported for Reelay.

We feed Reelay the same JSON traces produced by RyBench's schema-guided generators, covering random and fixed generation.
We check the resulting traces against three properties from the Timescales suite -- \textit{Bounded Response Globally}, \textit{Bounded Absence After q}, and \textit{Bounded Response Between q and r} -- chosen to show the range of formula complexity within the suite.
We instantiate Reelay's discrete-time monitor for each property, using the past-time formulation of each property as provided by the Timescales benchmark itself, since Reelay's specification language is past-time only.
An alternative technique -- evaluating a mirrored formula over a time-reversed trace -- exists for offline validation, but requires the complete trace in advance, so it does not apply here.
Incoming JSON is parsed with RapidJSON directly into a plain event structure, which Reelay consumes without constructing an intermediate JSON object, so this parsing cost is included in the measured latency.
The property column names only the Reelay specification used to check the trace; the underlying event schema and trace source are unchanged across properties, since neither generator depends on the guiding formula.
Table~\ref{tab:monitor-transport-latency} summarizes the results for the JSON encoding; the experiment includes only JSON because Reelay is evaluated as-is, varying only the transport layer.

\begin{table}[!b]\centering
\caption{Reelay monitoring latency per transport, measured with RyBench (JSON encoding, at 100\,kHz).}
\label{tab:monitor-transport-latency}
\vskip 0.3em
\footnotesize
\begin{tabular}{|l|l|l|r|r|r|}
\hline
\multirow{2}{*}{\textbf{generator}} & \multirow{2}{*}{\textbf{property}} & \multirow{2}{*}{\textbf{transport}} & \multicolumn{3}{c|}{\textbf{latency (ns)}} \\\cline{4-6}
 &  &  & \textbf{p50} & \textbf{p90} & \textbf{p99} \\\hline
\multirow{6}{*}{Schema-guided (random)} & \texttt{response\_globally} & iceoryx2 & \num{756} & \num{931} & \num{1480} \\\cline{2-6}
 & \texttt{response\_globally} & zenoh & \num{16888} & \num{20209} & \num{54582} \\\cline{2-6}
 & \texttt{absence\_after\_q} & iceoryx2 & \num{601} & \num{676} & \num{1051} \\\cline{2-6}
 & \texttt{absence\_after\_q} & zenoh & \num{18637} & \num{25533} & \num{60771} \\\cline{2-6}
 & \texttt{response\_between} & iceoryx2 & \num{820} & \num{966} & \num{1088} \\\cline{2-6}
 & \texttt{response\_between} & zenoh & \num{21501} & \num{28453} & \num{140898} \\\hline
\multirow{6}{*}{Fixed replay} & \texttt{response\_globally} & iceoryx2 & \num{728} & \num{888} & \num{1611} \\\cline{2-6}
 & \texttt{response\_globally} & zenoh & \num{21889} & \num{28598} & \num{143476} \\\cline{2-6}
 & \texttt{absence\_after\_q} & iceoryx2 & \num{685} & \num{811} & \num{1047} \\\cline{2-6}
 & \texttt{absence\_after\_q} & zenoh & \num{24573} & \num{31854} & \num{143749} \\\cline{2-6}
 & \texttt{response\_between} & iceoryx2 & \num{893} & \num{1076} & \num{1440} \\\cline{2-6}
 & \texttt{response\_between} & zenoh & \num{19515} & \num{27817} & \num{86751} \\\hline
\end{tabular}
\end{table}

Two observations follow from this experiment.
First, latency is dominated by the transport layer: Zenoh, which also supports shared memory but is run here over the network stack to make networking cost explicit, incurs roughly 20\,\textmu s of end-to-end latency even on a single host, since every message traverses system calls and the local network stack, whereas the shared-memory transport iceoryx2 stays below 1\,\textmu s at the median.
Second, the choice of formula has a visible effect on iceoryx2: p50 latencies differ by a few hundred nanoseconds across properties, with the most complex property, \emph{Bounded Response Between q and r}, incurring the highest latency under both generators. 
This ordering is consistent with the per-formula monitoring costs reported for Reelay itself~\cite{ulus2026reelay}, which measured the monitoring step in isolation, whereas our measurements are end-to-end; on the shared-memory transport, the monitoring cost thus remains observable within the total latency. 
On Zenoh, cross-property p50 spreads are larger, reaching several microseconds, consistent with the additional variance introduced by the network stack.
The open-loop rate of 100\,kHz was chosen to expose transport-layer costs under a controlled publishing load.
This rate is roughly 4\% of Zenoh's measured publishing capacity (Table~\ref{tab:perf-replay}), although the maximum sustainable monitoring rate and queueing behavior under saturation were not evaluated.

We also evaluate MonPoly using the same RyBench publishing pipeline.
Unlike Reelay, MonPoly runs as a separate process and consumes timestamped relational logs.
For this integration, we implemented a MonPoly-specific serializer that produces events in the required format and a subscriber-side adapter that forwards them through standard input and collects verdicts asynchronously from standard output.
The adapter associates each verdict with its timepoint and measures latency from the publication of the event that triggers the verdict to its collection.
This measurement includes the additional inter-process communication and verdict collection overhead.
To estimate the non-monitoring overhead, we replaced MonPoly with an echo process that immediately returns a verdict through the same adapter, obtaining median end-to-end latencies of 2.7--3.0\,\textmu s with iceoryx2 and 16.8--17.6\,\textmu s with Zenoh.
We use bounded future-time MFOTL formulations of the same three property categories, with time bounds of 10 steps.
The experiment uses both generators and transports at 100\,kHz, publishing 200,000 events per run, with three repetitions.
Table~\ref{tab:monpoly-transport-latency} summarizes the results.

\begin{table}[!b]\centering
\caption{MonPoly monitoring latency per transport, measured with RyBench (MonPoly log format at 100\,kHz).}
\label{tab:monpoly-transport-latency}
\vskip 0.3em
\footnotesize
\begin{tabular}{|l|l|l|r|r|r|}
\hline
\multirow{2}{*}{\textbf{generator}} & \multirow{2}{*}{\textbf{property}} & \multirow{2}{*}{\textbf{transport}} & \multicolumn{3}{c|}{\textbf{latency ($\mu$s)}} \\\cline{4-6}
 & & & \textbf{p50} & \textbf{p90} & \textbf{p99} \\\hline
\multirow{6}{*}{Schema-guided (random)} & \texttt{response\_globally} & iceoryx2 & \num{4.5} & \num{5.2} & \num{46.1} \\\cline{2-6}
 & \texttt{response\_globally} & zenoh & \num{18.3} & \num{21.2} & \num{697.6} \\\cline{2-6}
 & \texttt{absence\_after\_q} & iceoryx2 & \num{4.6} & \num{5.6} & \num{42.0} \\\cline{2-6}
 & \texttt{absence\_after\_q} & zenoh & \num{18.9} & \num{22.4} & \num{1024.3} \\\cline{2-6}
 & \texttt{response\_between} & iceoryx2 & \num{6.5} & \num{24.5} & \num{141.2} \\\cline{2-6}
 & \texttt{response\_between} & zenoh & \num{23.3} & \num{114.8} & \num{1643.1} \\\hline
\multirow{6}{*}{Fixed replay} & \texttt{response\_globally} & iceoryx2 & \num{4.8} & \num{5.3} & \num{44.5} \\\cline{2-6}
 & \texttt{response\_globally} & zenoh & \num{19.3} & \num{23.0} & \num{996.5} \\\cline{2-6}
 & \texttt{absence\_after\_q} & iceoryx2 & \num{4.6} & \num{5.7} & \num{42.1} \\\cline{2-6}
 & \texttt{absence\_after\_q} & zenoh & \num{19.8} & \num{24.0} & \num{1162.1} \\\cline{2-6}
 & \texttt{response\_between} & iceoryx2 & \num{6.2} & \num{19.6} & \num{108.1} \\\cline{2-6}
 & \texttt{response\_between} & zenoh & \num{23.9} & \num{137.4} & \num{1437.7} \\\hline
\end{tabular}
\end{table}

Two observations follow from this experiment.
First, the transport layer remains a significant contributor to end-to-end latency despite the additional process communication: iceoryx2 achieves median latencies of 4.5--6.5\,\textmu s, while Zenoh ranges from 18.3 to 23.9\,\textmu s.
Second, the choice of property also affects latency, with \emph{Bounded Response Between q and r} consistently showing higher median values under both generators.
The p99 latencies show greater variation, particularly with Zenoh, reaching over 1\,ms for some properties.
This experiment illustrates the integration of an external-process monitor alongside the in-process Reelay integration.
The results are not intended as a direct performance comparison between the two tools, since their specifications and integration mechanisms differ, but they provide insights into potential performance bottlenecks in their respective monitoring pipelines.

\section{Conclusion and Future Work}
\label{sec:conclusion}

This paper presented RyBench, a benchmarking framework for RV tools operating over pub/sub networks.
RyBench keeps event generation, serialization, and transport as independently resolvable components, each defined behind an abstract interface and selected through build-time feature flags and run-time configuration.
This separation keeps the design modular, making it easy to add new generators, serializers, and transports.
It also allows the same components to support both online generation, which publishes events live over the pub/sub network, and offline generation, which persists them to a file for later replay.
As a result, an RV tool's transport and parsing overhead can be benchmarked directly and in isolation from the monitoring procedure itself.

Our application to Reelay demonstrates the effect of transport choice on end-to-end monitoring latency, with Zenoh exhibiting median latencies more than an order of magnitude higher than iceoryx2.
The MonPoly experiment further illustrates how RyBench can be integrated with an external-process RV tool using a tool-specific serializer and a subscriber-side adapter.
Our experiments further show that RyBench retains high throughput despite its modular design, sustaining millions of events per second for random and fixed generation.
Because schemas and the selection among available components are resolved at run time, RyBench can be configured for different message schemas, properties, and transports without modifying the framework itself, making it practical to reuse across different RV tools and target systems.

RyBench's random and fixed generators, described in Section~\ref{sec:arch-generation}, cover functional testing and reproducible replay, respectively.
The property-guided generator, outlined in the same section, would drive generation from a temporal-logic property rather than a fixed schema, targeting arbitrary future-time properties.
It would generate traces by tracking the residual obligations of the property along its syntax tree, producing each event through a lightweight traversal instead of a full satisfiability check, trading the completeness guarantees of solver-based approaches for speed.
This trade-off means that the proposed generator may fail to produce a trace even for a satisfiable property and does not guarantee representative coverage of satisfying traces.
Implementing and evaluating this generator is left for future work.
Such property-guided traces let RyBench stress-test an RV tool's monitoring procedure itself, not just its transport and parsing pipeline.
Latency can therefore be measured as a function of how close a trace runs to a verdict, since trace content is shaped by the guiding property rather than being arbitrary.
The same mechanism can further be extended to produce violating traces paired with a verdict oracle, enabling automated correctness testing of RV tools.

RyBench can be used for cross-tool comparisons by evaluating semantically equivalent properties under common workloads and publishing conditions.
However, differences in specification languages, property translations, and integration mechanisms (e.g., in-process versus inter-process) can affect the measured performance independently of the monitoring algorithm and should be accounted for when interpreting the results.

The current evaluation focuses on the one-to-one topology on a single host.
Although autonomous systems may be distributed across hosts, individual subsystems can still contain multiple processes on the same host, making shared-memory transport relevant for local monitoring.
More complex deployments could combine such local monitors with an orchestration and verdict-aggregation component, introducing additional communication overhead.
The additional topologies described in Section~\ref{sec:arch-transport} cover settings with multiple publishers or subscribers that could be used to study these configurations.
Evaluating these topologies is left for future work to determine whether the throughput and latency characteristics observed here hold as communication complexity increases.

Deploying across physically separate hosts would extend this further: our experiments run Zenoh over the network stack of a single host, which makes networking cost explicit but does not expose effects such as jitter and variable transmission delay that only arise once hosts are physically separated.
Latency across hosts can only be measured relative to a shared clock, and any error in synchronizing that clock is indistinguishable from latency in the measurement.
This is not a minor concern for RyBench specifically, since our single-host measurements already report sub-microsecond medians for the shared-memory transport; once hosts are physically separated, even small synchronization errors could dominate the measurement.

\section*{Data and Code Availability}
RyBench, together with the benchmark configurations, the schemas of experiment setups, and the scripts used to produce Tables~\ref{tab:perf-random}--\ref{tab:monpoly-transport-latency}, is publicly available at \url{https://github.com/bounverif/rybench}.

\section*{AI Usage}

An AI assistant (Claude, Anthropic) was used during the preparation of this manuscript to help draft and revise prose -- improving grammar, sentence flow, and the organization of paragraphs and sections -- and to assist with \LaTeX{} formatting, including figure code.
All research contributions reported in this paper, including the design of RyBench, the property-guided generation algorithm, and the experimental results, are the authors' own work; the AI assistant was not used to generate research ideas, algorithms, or results, and all AI-assisted text was reviewed and edited by the authors.
 
\bibliographystyle{eptcs}
\bibliography{paper}
\end{document}